\documentclass[graybox]{svmult}

\makeatletter
\renewcommand{\cleardoublepage}{\clearpage}
\renewcommand{\chapter}{%
  \clearpage
  \thispagestyle{plain}%
  \global\@topnum\z@
  \@afterindenttrue
  \secdef\@chapter\@schapter
}
\@twosidefalse
\makeatother
\usepackage{type1cm}        
\usepackage{makeidx}         
\usepackage{graphicx}        
\usepackage{multicol}        
\usepackage[bottom]{footmisc}

\usepackage{newtxtext}       %
\usepackage[varvw]{newtxmath}       

\makeindex             

\begin{document}

\title*{Fundamentals of Quantum Machine Learning and Robustness}
\author{Lirandë Pira and\\ Patrick Rebentrost}
\institute{Lirandë Pira\orcidID{0000-0002-6305-1150} \at Centre for Quantum Technologies, National University of Singapore, \email{lpira@nus.edu.sg}
\and Patrick Rebentrost\orcidID{0000-0002-6728-8163} \at Centre for Quantum Technologies, National University of Singapore \email{cqtfpr@nus.edu.sg}}

\maketitle

\abstract{Quantum machine learning (QML) sits at the intersection of quantum computing and classical machine learning, offering the prospect of new computational paradigms and advantages for processing complex data. This chapter introduces the fundamentals of QML for readers from both communities, establishing a shared conceptual foundation. We connect the worst-case, adversarial perspective from theoretical computer science with the physical principles of quantum systems, highlighting how superposition, entanglement, and measurement collapse influence learning and robustness. Special attention is given to adversarial robustness, understood as the ability of QML models to resist inputs designed to cause failure. We motivate the study of QML in adversarial settings, outlining distinctions between classical and quantum data and computations when the adversary is a core element. This chapter serves as a starting point to  adversarial and robust quantum machine learning in subsequent chapters.}

\section{Introduction}\label{sec:introduction}
Quantum machine learning (QML) covers the intersection of quantum computing and artificial intelligence \cite{schuld2021machine,biamonte2017quantum}. Machine learning is able to extract patterns from large datasets and improve performance with experience. Machine learning  achieves success in applications such as image and speech recognition, medical diagnosis, financial forecasting, and autonomous systems \cite{lecun2015deep,goodfellow2016deep}. Quantum computing is a new paradigm of computing based on the principles of quantum mechanics \cite{benioff1980computer,nielsen2010quantum}. These principles characterized in superposition, entanglement and quantum interference are used to process information in new ways. Quantum algorithms have demonstrated the potential to solve certain problems exponentially faster than their classical counterparts. A prominent example is Shor’s algorithm for integer factorization, which can factor large numbers in polynomial time, thereby posing a theoretical threat to classical cryptographic systems \cite{shor1997polynomial}. This breakthrough, among others, inspires interest in identifying domains where quantum speedups could be achieved \cite{montanaro2016quantum}.

Machine learning relies on linear algebra calculations and optimizations. Quantum computing is about processing vectors in high-dimensional vector spaces. Combining these ideas has led to explorations of quantum subroutines within machine learning pipelines, particularly in the context of model training and inference. Notably, QML is motivated by potential quantum advantages, either in speed, representational capacity, or data efficiency, for learning tasks that are classically difficult. Yet, realizing these advantages in practice requires more than computational power. Increasing system complexity raises questions of reliability, robustness, and interpretability. The same features that define quantum systems, such as superposition, entanglement, and interference, also make them highly sensitive to noise and perturbations. Robustness therefore becomes a central concern in the development of QML.

A model that performs well under ideal conditions but fails under small perturbations is of limited practical use, especially in sensitive domains. This concern is amplified by the fact that quantum systems are themselves noisy and sensitive to small errors. If we hope to deploy QML systems in real-world settings, they must be able to handle uncertain environments and potential manipulation. This practical concern motivates the growing field of quantum machine learning that explicitly addresses robustness \cite{west2023towards}.

This chapter aims to establish the foundational context for QML, offering readers a conceptual framework that connects classical machine learning with principles of quantum computation. An emphasis will be placed on adversarial learning as a paradigm within learning systems, while considering the broader approaches to robustness. By doing so, the goal of this chapter is to serve as a starting point for exploring adversarial quantum learning and robustness in QML.

\section{Learning and Computation}
\subsection{Role of Machine Learning in Modern Computation}
Machine learning has influenced both the practical and theoretical landscapes of modern computation \cite{bishop2006pattern,hastie2009elements}. Beyond large-scale data-driven industrial applications, it has introduced computational paradigms for algorithmic thinking.

On a theoretical level, machine learning establishes a connection between statistics and computer science. It formalizes the process of learning as an optimization problem over hypothesis spaces with a focus on inference and generalization \cite{valiant1984theory}. This connection has expanded the scope of computation beyond deterministic procedures, including probabilistic reasoning and approximation methods as central components. Additionally, machine learning theory explores questions of model complexity and expressivity, investigating how models generalize from finite samples, and the computational resources required to learn certain classes of functions. Such insights have led to the development of algorithmic frameworks that accommodate uncertainty, and interaction with evolving environments. For example, online and reinforcement learning models learn and evolve over time, going beyond the static input-output nature of algorithms. Robustness and stability, broadly defined as how computational models withstand noise, adversarial inputs, and uncertainty, are essential theoretical concerns. Addressing these challenges has led to refined complexity measures and formal definitions, including average-case and data-dependent analyses.

Machine learning has grown in waves, each clearing a practical hurdle and setting up the next. Early ideas on the perceptron, formal notions of what can be learned, and effective backpropagation, showed that adjusting many parameters could still yield reliable generalization \cite{rosenblatt1958perceptron,vapnik1971vc,valiant1984theory,rumelhart1986learning}. The 1990s brought support vector machines, boosting, and kernels, giving practitioners tools that worked well out of the box \cite{cortes1995support,freund1997decision,scholkopf2002learning}. Around the 2000s, probabilistic graphical models and faster training routines made it easier to encode structure and uncertainty \cite{robbins1951stochastic,nesterov1983method,koller2009probabilistic}. A turning point followed with unsupervised layer-wise pretraining and large convolutional networks \cite{hinton2006fast,krizhevsky2012imagenet}, which helped push widespread use of GPU (graphical processing unit) computation. Stability techniques such as normalization, residual links and adaptive optimizers, made very deep models practical \cite{ioffe2015batch,he2016deep,kingma2015adam}. Self-attention and the Transformer \cite{vaswani2017attention} simplified handling long-range context, leading to larger models \cite{kaplan2020scaling,belkin2019reconciling}. Most recently, foundation and diffusion models \cite{brown2020language,ho2020denoising} combine these advances with broader pretraining and emergent in-context behavior. Each stage thus reinterprets learning: from formal feasibility, to deep representation, to broad flexible pretrained systems. This history underlies today’s robustness questions and informs how we think about future learning pipelines, including quantum ones.

In summary, machine learning has transformed what computers can do, all the while expanding how we conceive and define computation itself. Algorithms are no longer static input–output mappings but adaptive optimization processes whose reliability depends on capacity control, training dynamics, and statistical validation. In this book, these theoretical foundations provide a natural stepping stone to quantum settings. Quantum algorithms introduce different mechanisms for information processing that promise new approaches to learning tasks. Beyond quantum-classical heuristics, the interplay between quantum mechanics and learning theory is a rapidly developing area within QML research \cite{arunachalam2017survey}. Below, we outline the broad motivation and current state of quantum computation as the hardware platform on which QML is built.

\subsection{Present Day Quantum Computation}
Quantum computing departs from classical computation by quantum mechanical principles such as superposition, entanglement, and unitarity \cite{nielsen2010quantum,feynman1982simulating}. The fundamental unit of information is the qubit, which unlike a classical bit, can encode a continuum of states between $0$ and $1$. Multiple qubits form a quantum register whose state is described by a vector in a high-dimensional Hilbert space, with dimension exponential in the number of qubits. Quantum states evolve through the application of unitary transformations, which are reversible operations that preserve the norm of the state vector. In practice, these transformations are implemented as sequences of quantum gates acting on single or pairs of qubits. Computation proceeds by preparing an initial quantum state, applying a meaningful gate sequence, such as a quantum algorithm, and finally measuring the result in the computational basis to extract classical outcomes.

Quantum computing hardware becomes increasingly capable. Platforms range from superconducting qubits to trapped ions and photonics. Current quantum hardware, known as noisy intermediate-scale quantum (NISQ) near-term devices, provides access to tens to hundreds of physical qubits \cite{preskill2018quantum}. NISQ systems are limited by decoherence, gate infidelity, and noise accumulation, which restrict the depth of circuits and algorithmic complexity that can be executed. Despite these constraints, these system sizes offer a testbed for exploring certain quantum algorithms, quantum control techniques, and hybrid quantum-classical routines.

It is worth highlighting that the main driver of progress in quantum computing is the pursuit of quantum advantage, the point at which a quantum device performs a computational task demonstrably better than any classical counterpart \cite{harrow2017quantum,arute2019quantum,zhong2020quantum}. This could mean faster runtime, lower memory requirements, or the ability to solve problems that are difficult for classical algorithms. While early demonstrations have focused on narrow, contrived tasks, the broader goal is to realize advantage in meaningful applications, including optimization, simulation, and machine learning. This is where fault-tolerant quantum computing comes into the picture. Fault-tolerant quantum computation is the paradigm that refers to large-scale quantum computers with meaningful processing capacities, capable of achieving quantum advantage \cite{shor1997faulttolerantquantumcomputation}, as opposed to limited NISQ devices. Achieving fault tolerance relies on quantum error correction, which encodes logical qubits into many physical ones to protect against decoherence. 

Before full fault tolerance is available, much of the focus in the NISQ era is on the near-term algorithms that can run on noisy hardware. Variational quantum algorithms, such as the variational quantum eigensolver \cite{peruzzo2014variational} and quantum approximate optimization algorithm \cite{farhi2014quantumapproximate} are designed to be resilient to some degree of noise, using classical optimizers to tune quantum circuits. Within QML, such hybrid approaches are given by quantum neural networks, which combine quantum state evolution with classical gradient-based training.

Currently the hardware platforms share common bottlenecks such as gate fidelity, readout precision, and qubit stability, among others. It is widely accepted that achieving meaningful quantum advantage in practical tasks will require both an increase in the number and quality of qubits, i.e., in improved coherence times, reduction of gate errors, and the development of architectures that support error correction at scale. At the same time, on the theoretical front, it remains equally important to understand what can actually be learned, computed, or inferred with limited quantum resources. This understanding plays a significant role in guiding the development of quantum algorithms today.

\section{Understanding Robustness in Machine Learning}
Computer science fundamentally approaches problems from a worst-case perspective. Unlike many fields that study typical or average scenarios, theoretical computer science asks: \textit{How hard is the problem in the most difficult or adversarial instance?} This worst-case mindset is central to several pillars of the discipline. For instance, algorithmic complexity is defined by the hardest possible inputs an algorithm might face. A problem is deemed ``hard'' not because it is difficult on average, but because it is difficult for some worst-case instances. Similarly, security and cryptography are built on the assumption that adversaries can be highly powerful and strategic, so systems must remain secure even against the most damaging attacks. In this way, the worst-case analysis takes adversarial robustness as a fundamental concern in algorithm design.

Robustness, in the learning setting, asks how performance degrades when the real-world setting departs, slightly or substantially, from idealized modeling assumptions. This can occur when inputs are perturbed, distributions shift, labels are corrupted, data are strategically selected, or the model itself is probed for information leakage. Historically, robustness-oriented thinking in learning predates modern deep systems. Robust statistics introduced the idea that estimators should retain stability under a fraction of arbitrary contamination \cite{huber1964robust,tukey1960survey}. As machine learning met adversarial behavior in applied settings, early spam filtering and intrusion detection research exposed the strategic nature of evasion and feature manipulation \cite{dalvi2004adversarial,lowd2005adversarial}. Domain adaptation theory subsequently clarified how one can transfer performance guarantees across related but shifted distributions \cite{ben2010theory}. Distributionally robust optimization crystallized these themes in min-max formulations over uncertainty sets, anticipating the empirical saddle-point structure of contemporary adversarial training \cite{ben2013robust,duchi2016statistics,namkoong2017variance}.

\begin{figure}[b]
\centering
\includegraphics[width=0.89\linewidth]{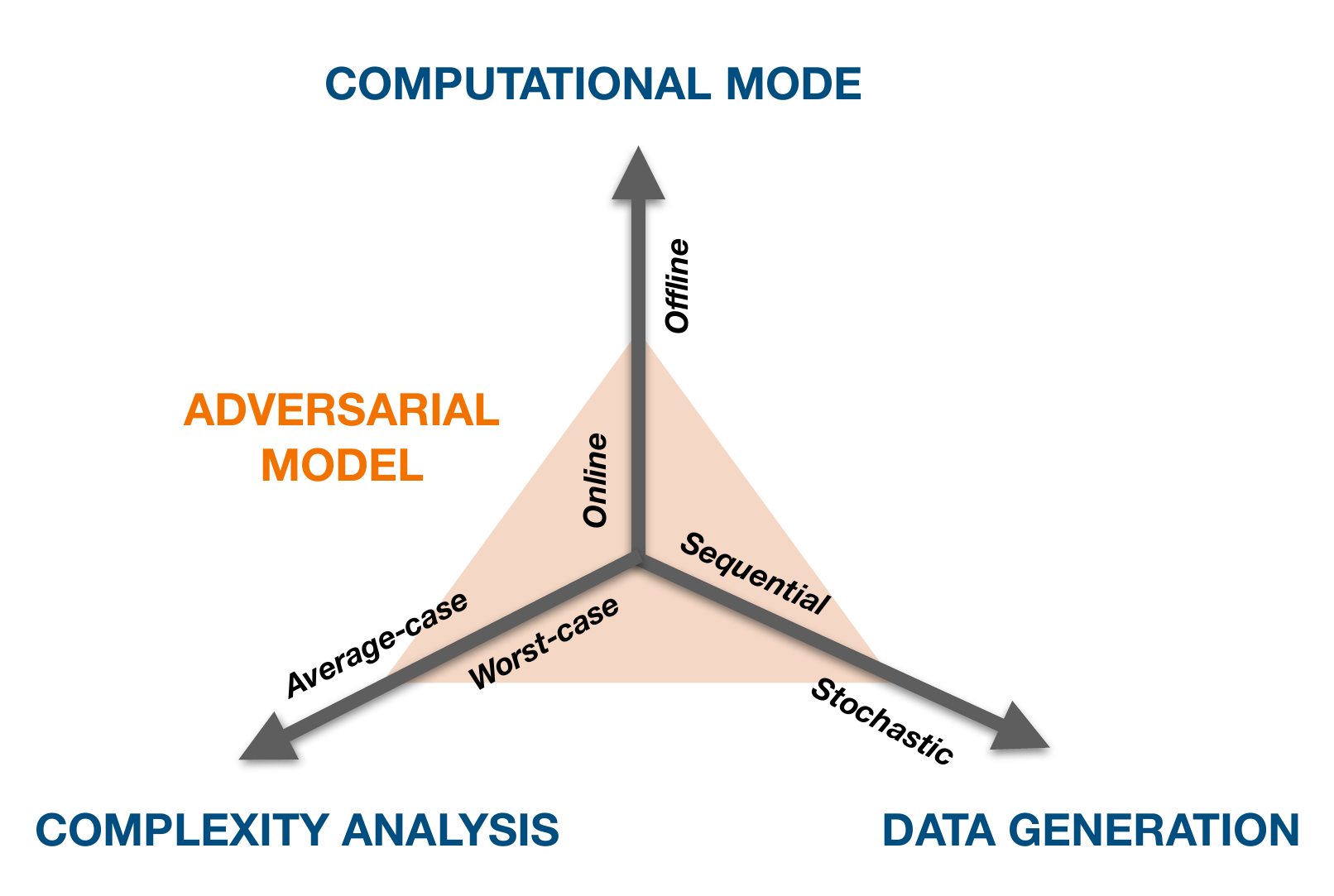}
\caption{A conceptual map positions learning settings along data generation, computation mode, and complexity analysis dimensions, showing where adversarial models naturally arise (such as the shaded area in orange) and why robustness considerations may vary across this spectrum.}
\label{fig:modes}
\end{figure}

In understanding learning and computation under uncertainty, it is helpful to view different frameworks along a spectrum defined by three axes: the nature of complexity analysis, the computational mode, and how data is generated or accessed, as captured in Fig. \ref{fig:modes}. Complexity theory traditionally begins with worst-case analysis, asking how algorithms behave under the hardest possible inputs. Average-case complexity and probabilistic analysis relax this analysis by assuming distributions over inputs. Similarly, computational models range from online settings, where decisions are made sequentially without future knowledge, to offline models that assume full data access. Finally, data generation spans from adversarial or deterministic sequences to stochastic models, such as i.i.d. sampling. As an example, multi-armed bandits illustrate these concepts well, with stochastic bandits representing average-case learning under uncertainty, and adversarial bandits capturing worst-case sequential interaction. Online learning, adversarial machine learning, and robust optimization sit toward the worst-case, sequential, adaptive end of the spectrum, while classical supervised learning lives on the average-case, offline, i.i.d. side.

And finally, in real-world scenarios, drawing sharp boundaries between concepts is rarely natural. Theoretical distinctions such as worst-case versus average-case analysis, or online versus offline models, are useful abstractions, but they rarely capture the nuances of practical problems. In reality, there’s a spectrum. It is worth noting that worst-case complexity bounds are also problem-specific. Thus, the structure of the problem changes the nature of worst-case analysis. As settings move from full information (such as offline) to partial, sequential, or reactive settings (such as online), worst-case analysis becomes both more relevant and more subtle. In adversarial contexts, these distinctions become important when designing robust algorithms, including quantum ones. This perspective helps clarify how different fields relate, and why worst-case thinking becomes essential in adversarial and high-stakes environments, which is often referred to in this book.

In models of online algorithms and adversarial learning, inputs may be chosen sequentially by an adversary aiming to cause the worst possible outcome. A compelling example is when the adversary may select the very first data point in a way that maximally misguides the learning algorithm. This perspective reflects the reality of many practical settings. These adversarial environments are exemplified in security, finance, and autonomous systems, where worst-case scenarios must be handled robustly.

\section{Core Concepts in Quantum Machine Learning}
\subsection{Motivation}
Machine learning is fundamentally about using data to make predictions or optimize decisions. The typical approach involves training a model to minimize a given loss function, which measures how far its predictions deviate from expected outcomes. While modern machine learning systems, especially deep learning models, can be powerful, they often come with high computational and data requirements. Real-world practical deployment of machine learning systems often requires large and scalable models, which in turn increases the cost of training and inference as the models scale. This fact, coupled with the emergence and promises of quantum computing, has led to growing interest in whether quantum computing could help. The motivation for QML nowadays goes beyond speed, although that is often cited. Equally appealing is the idea that quantum systems can represent and manipulate information in ways that classical systems cannot, thereby capturing structural representations intractable classically. For instance, a quantum system with $n$ qubits has access to a Hilbert space of size $2^n$, which allows, at least in principle, compact representations of very complex functions or data distributions. This might enable more efficient algorithms for certain tasks such as kernel evaluation, sampling, or optimization, which are common bottlenecks in classical machine learning. Unsurprisingly, these computational bottlenecks became some of the earliest targets for exploration in QML research. These early explorations naturally gave rise to a broader conceptual mapping of QML, clarifying the different ways quantum and classical components can be combined in learning systems.

\begin{figure}[b]
\centering
\includegraphics[width=0.89\linewidth]{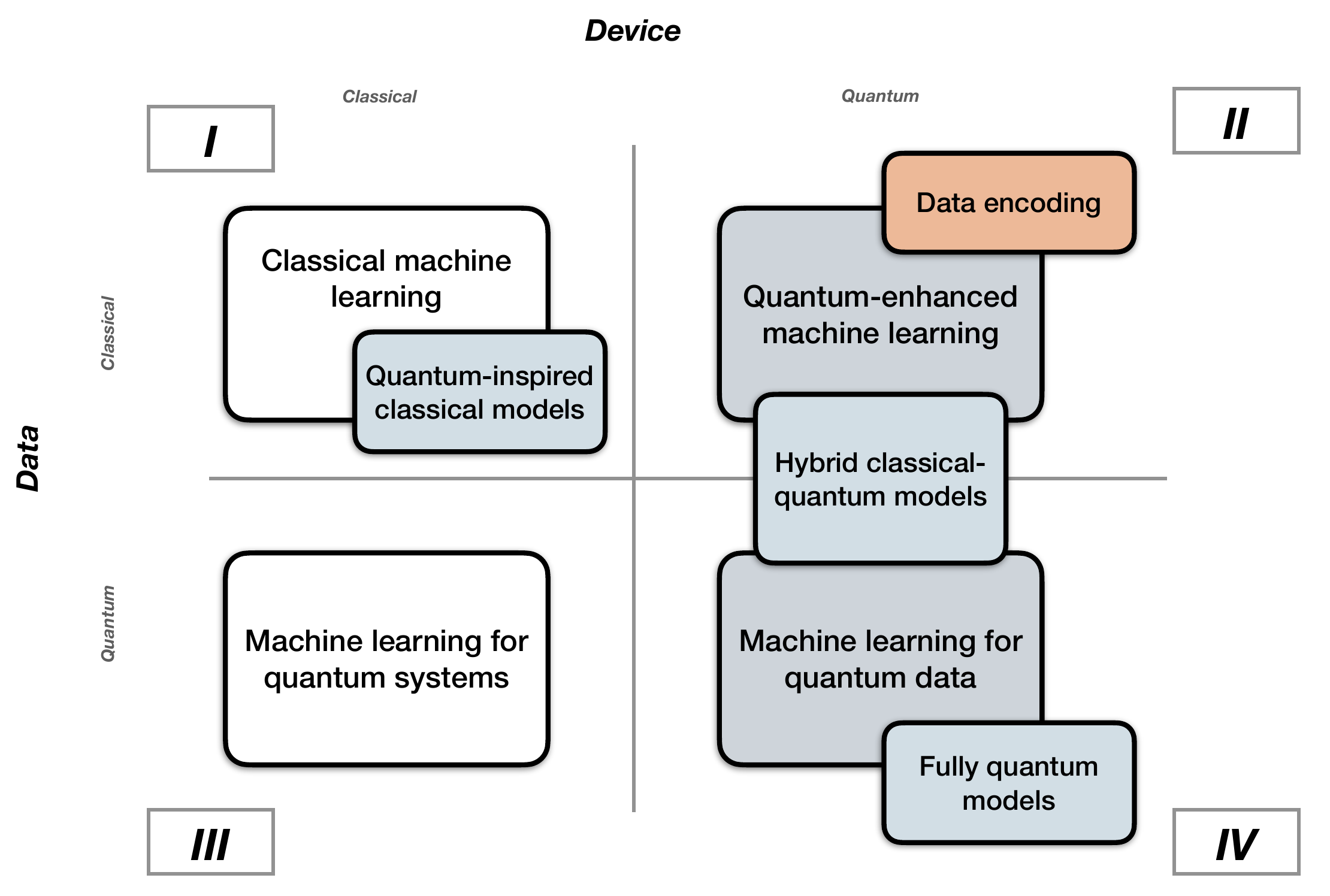}
\caption{Taxonomy of quantum machine learning approaches. This diagram maps QML methods along two axes: the nature of the data (classical or quantum) and the computational resource (classical or quantum). The top-right quadrant corresponds to quantum-enhanced machine learning, where quantum models are applied to classical data. The bottom-right represents machine learning for quantum data, where quantum systems or outputs are analyzed. Classical machine learning occupies the top-left along with quantum-inspired classical models, while the bottom-left includes classical models used to support quantum computing tasks such as control and error correction.}
\label{fig:quadrants}
\end{figure}

\subsection{Paradigms}
In light of these motivations, QML research has converged on several broad paradigms. The first main approach is \textit{quantum-enhanced machine learning} which covers scenarios where quantum computing is used to improve or accelerate standard machine learning tasks. Here, data may be classical, but quantum devices are employed to perform operations such as kernel evaluation, matrix inversion, optimization, or sampling. The second one can be viewed as \textit{machine learning for quantum data}. Here, the learning task itself involves quantum states as input or output. Examples include learning properties of quantum systems, extracting information from quantum experiments, or modeling outcomes of quantum processes. These distinctions are not mutually exclusive, and similar to broader QML, the boundaries are often fluid. Beyond the two broad views discussed above, there are at least two other perspectives worth highlighting. First, there is \textit{machine learning for quantum systems}, where classical learning methods are used to assist in the development and control of quantum devices themselves. This includes tasks such as quantum error correction or pulse shaping. In these settings, the learning aspect supports the underlying challenges on quantum hardware. Secondly, \textit{quantum-inspired classical models}, carries ideas or techniques from quantum computing, such as tensor networks or Hilbert space embeddings, for purely classical algorithms \cite{tang2019quantum}. This approach aims to replicate some of the efficiency or structure of quantum systems using classical resources, without requiring access to a quantum device.

More generally, this taxonomy and broad categorization helps understand what parts of the QML landscape are physically realizable today, which are targets for fault-tolerant quantum computation, and which lie outside hardware constraints but still inform algorithmic design. The nature of the input data, whether classical or quantum, dictates what computational resources are needed and what types of learning models are feasible. 

This taxonomy, summarized in Fig. \ref{fig:quadrants}, organizes QML methods along two orthogonal dimensions: the nature of the data and the computational resource or model. The following two subsections explore these axes in more detail, beginning with the data dimension.

\subsection{Data Representation and Encoding}
In conventional machine learning, data points are typically structured arrays over real or discrete fields, for example, vectors, sequences, or images with clear physical interpretation. Quantum data, by contrast, consists of pure or mixed quantum states in a Hilbert space. Such states may exhibit superposition and entanglement, properties with no straightforward classical analogues, and are subject to constraints such as the no-cloning theorem and the collapse of the wavefunction upon measurement.

This difference in representation is both a challenge and an opportunity. Quantum data can encode correlations, interference patterns, and high-dimensional structures that classical systems struggle to represent efficiently. However, it is also fragile: extracting information risks disturbing or destroying the state, and classical simulation of large quantum systems quickly becomes computationally prohibitive.

When working with classical datasets on a quantum computer, an additional step is required: encoding the classical data into quantum states \cite{schuld2021machine,larose2020robust}. Several approaches exist. In \emph{basis encoding}, each classical bitstring is mapped directly to a computational basis state $\lvert x \rangle$. \emph{Amplitude encoding} embeds a normalized real or complex vector into the amplitudes of a quantum state, allowing $n$ qubits to represent $2^n$ features compactly. \emph{Angle} or \emph{rotation encoding} instead maps classical values to the rotation angles of single-qubit gates, a technique common in variational circuits. More elaborate \emph{hybrid schemes} combine these strategies or exploit problem-specific symmetries to optimize resource usage. A refinement particularly relevant for rotation-based schemes is \emph{data re-uploading} method, which repeatedly inserts the input features across layers to hold qubit count fixed \cite{perezsalinas2020data}.

This raises a natural question: to what extent do speedups depend on the nature of the data itself? Research on classical–quantum separations suggests that certain advantages may hinge critically on whether the input is quantum or classical \cite{huang2021power}. This connects directly to robustness: a computational advantage that vanishes under small perturbations in the data is, in practice, no advantage at all.

Yet, while the way information is encoded determines what structure is accessible to a learner, the choice of architecture governs how that structure is exploited. Together, these observations suggest that questions of data representation cannot be separated from questions of model design. It is therefore natural to turn next to model architectures, where different computational assumptions and hardware constraints give rise to distinct approaches to quantum learning.

\subsection{Model Architectures}
Work on model architectures divides broadly into two regimes. A first line of QML research targets the fully fault-tolerant regime, where quantum error correction enables deep, precise circuits and algorithms with formal asymptotic guarantees, albeit often under strong assumptions about data access and state preparation \cite{aaronson2015read}. At a high level, these models frequently draw on quantum linear algebra primitives, such as procedures for solving linear systems, estimating eigenvalues, or transforming singular values, that form the backbone of many classical machine learning methods. Early examples include algorithms such as Harrow-Hassidim-Lloyd (HHL) for linear systems \cite{harrow2009quantum} and related routines for tasks such as principal component analysis \cite{lloyd2014quantum} and kernel methods \cite{rebentrost2014quantum}. More recently, the framework of quantum singular value transformation (QSVT) \cite{gilyen2019quantum} has emerged as a unifying abstraction for these linear-algebraic building blocks. QSVT makes it possible to implement polynomial transformations of singular values with optimal query complexity, providing a clean theoretical foundation for many large-scale QML algorithms that are actively reshaped \cite{ivashkov2024qkanquantumkolmogorovarnoldnetworks}.

That said, most near-term quantum machine learning research does not aim for full quantum advantage in the theoretical sense. Instead, it focuses on hybrid models that use quantum circuits for parts of the computation, together with classical optimization or pre-processing. These efforts are broadly characterized as variational quantum algorithms \cite{cerezo2021variational,benedetti2019parameterized}, which are explicitly designed for noisy NISQ devices.

Within the broader variational algorithms, a commonly studied model architecture are quantum neural networks, which attempt to mimic the behavior of classical neural network by introducing a layered architecture. These models implement parameterized quantum circuits in order to tune the learnable parameters. While the model itself may include quantum components such as the parameterized circuits, the parameters are often updated using classical optimization algorithms. Training typically involves a hybrid loop where a quantum processor evaluates the circuit and a classical optimizer updates parameters.

A central technical question is how to obtain gradients of the loss with respect to circuit parameters. Unlike classical neural networks, backpropagation cannot be applied directly due to the nature of quantum measurements. Instead, the most common approach is the parameter-shift rule \cite{mitarai2018quantum,schuld2019evaluating}, which exploits the trigonometric structure of many quantum gates to compute exact gradients by evaluating the circuit at two shifted parameter values. This makes it possible to apply gradient-based optimizers such as stochastic gradient descent in a quantum–classical loop. However, while the parameter-shift rule enables proof-of-concept training, it does not fully address optimization and scalability challenges, which remain ongoing open questions.

In parallel, another practical line of work centers on quantum kernel methods \cite{havlicek2019supervised,schuld2019quantum}. Here, quantum circuits are used not as trainable models but as feature maps, embedding classical data into high-dimensional Hilbert spaces where classical kernel methods can operate.

One trend that currently stands out is the push to identify where quantum machine learning is not only viable but genuinely useful, often while drawing comparisons with classical counterparts \cite{bowles2024betterclassicalsubtleart}. To that end, research focus relies on a growing set of toy problems and benchmarks such as low-dimensional datasets, small-scale classification tasks, or quantum-native problems such as state discrimination, to test model behavior and scaling. There is particular interest in applications to quantum chemistry, where the structure of the problem maps naturally onto quantum systems, and in problems where sampling from high-dimensional distributions is computationally expensive \cite{cao2019quantum}. As these models grow in complexity, questions about their expressivity and capacity become increasingly important. Inevitably, understanding what these models can represent, and how that compares to classical baselines, remain central to evaluating their potential in practice. These benchmark settings often expose both the strengths and current limitations of QML approaches.

In summary, QML exists at the intersection of two distinct paradigms: quantum computation and statistical learning. Each brings its own strengths, and together they open up opportunities that classical approaches alone struggle to match. At the same time, QML inherits limitations from both fields, along with a set of unique obstacles that we are only beginning to understand. Below, we outline the potential benefits of quantum approaches to learning, while also being candid about the barriers ahead.

\section{Opportunities and Challenges}
\subsection{Algorithmic Speedups and Representational Capacity}
One of the most anticipated advantages of QML is the potential for algorithmic speedups. Certain quantum algorithms offer runtime improvements for linear algebra subroutines under constraints, which are central to many machine learning models \cite{morales2025quantumlinearsolverssurvey}. As mentioned previously, techniques such as the HHL algorithm for solving linear systems or the QSVT framework, suggest that quantum models might reduce the computational complexity of training and inference in certain settings. While these speedups are typically proven under idealized assumptions, they motivate the search for practical QML models that can offer meaningful improvements on real-world problems.

Beyond raw speed, quantum representations may allow for richer encodings of data. Quantum states naturally capture high-dimensional structure through entanglement and superposition. This can enable feature spaces that are exponentially large in the number of qubits, allowing quantum models to access complex decision boundaries that would be difficult to learn classically. Quantum feature maps have been proposed in this context, aiming to embed classical data into Hilbert spaces that better expose class structure \cite{schuld2019quantum,havlicek2019supervised}. Another promising direction is unsupervised learning. Classical unsupervised methods, such as clustering or dimensionality reduction, often struggle in high-dimensional or noisy settings. Quantum models, especially those using variational circuits or quantum annealing, may offer new ways to explore data structure, optimize clustering objectives, or sample from complex distributions. Explorations into applications are emerging across domains. In quantum chemistry and materials science, QML techniques are explored to model molecular properties more efficiently \cite{cao2019quantum}. In finance, quantum algorithms have been applied to examples of portfolio optimization and risk modeling \cite{orus2019quantum}. However, as we will discuss in the following, these opportunities must be viewed alongside serious present-day challenges.

\subsection{Hardware and Optimization Limitations}
Currently, quantum hardware remains noisy and error-prone with decoherence, imperfect gate fidelities, and limited qubit connectivity. Many QML proposals assume ideal unitary operations that are far from what current hardware can support. Another limitation arises in data encoding. As we noted, for a quantum algorithm to process classical data, the data must first be embedded into quantum states. This encoding step is often costly and may offset any potential quantum advantage. Furthermore, the amount of data that can be encoded is constrained by the number of available qubits, placing practical limits on input size and resolution. Optimization is also a major bottleneck. Variational quantum algorithms, which are central to many QML architectures, suffer from barren plateaus, or regions of the parameter space where gradients vanish exponentially with system size \cite{mcclean2018barren,larocca2025barren}. This makes training deep or expressive models difficult. Hybrid optimization strategies are being actively explored, however presently, scalable and stable training remains an open problem. There is also a lack of standardized datasets and benchmarks that are native to the quantum domain. Much of the current research relies on synthetic datasets or repurposed classical data to benchmark toy examples. All of these constraints limit the ability to evaluate models rigorously and meaningfully \cite{bowles2024betterclassicalsubtleart}. The same dataset constraints can be further noted in adversarial contexts where fine-grained control over the input distribution is often required.

Finally, the theoretical foundations of QML are still developing with questions around expressivity, generalization, and learnability \cite{caro2022generalizing}. While some results provide bounds on what quantum models can represent or learn, these are often limited in scope or dependent on strong assumptions. In summary, QML holds considerable promise, particularly for tasks that involve complex data structures or adversarial interactions. However, realizing that promise will require overcoming substantial technical hurdles, both in hardware and in algorithm design. The remainder of this book addresses one of the most important open questions in this field: how quantum models behave under adversarial conditions, and whether their unique properties offer any advantage in terms of robustness, or more conceptually, what further unique hurdles do QML systems bring the broader adversarial machine learning.

\section{Robustness and Quantum Adversarial Learning}
\begin{figure}[b]
\centering
\includegraphics[width=0.59\linewidth]{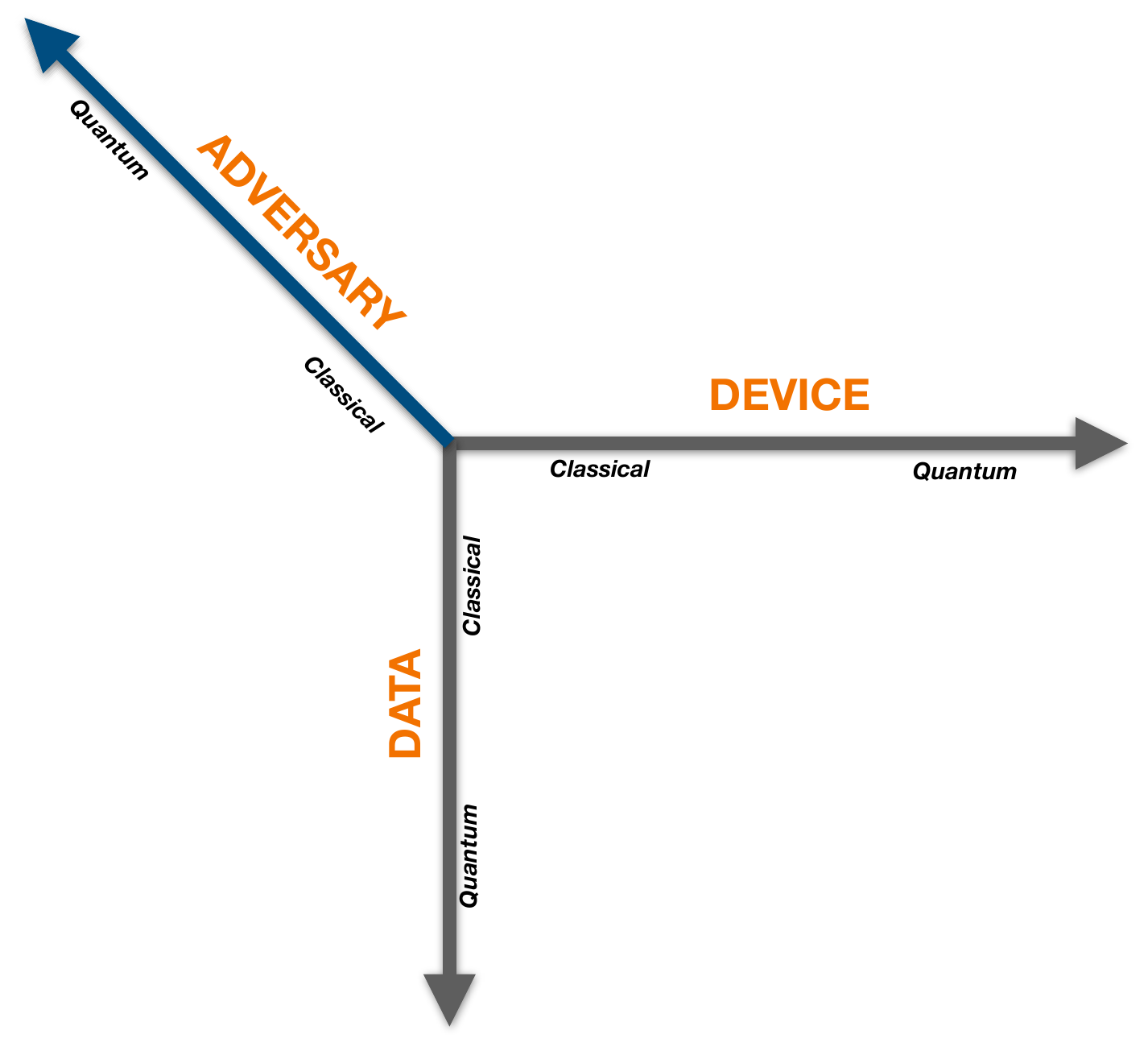}
\caption{Extension of the quantum-classical data-device paradigm to adversarial settings. The figure illustrates classical vs. quantum distinctions along the well-known data (vertical) and device (horizontal) axes, and introduces the adversary as a third component. This extension makes it possible to ask systematic questions such as: \textit{How does the nature of the adversary change when data or computation is quantum?} \textit{Which attack surfaces exist in each quadrant?} In this way, the framework helps map out where robustness challenges and opportunities arise in hybrid quantum–classical learning systems.}
\label{fig:dim3robustness}
\end{figure}
The motivation for robust learning systems extends into quantum learning \cite{west2023towards,lu2020quantum,ren2022experimental}. This book is motivated by a central hypothesis: quantum information and computation provide fundamentally new avenues for exploring robustness in machine learning. Furthermore, quantum systems exhibit properties such as state collapse upon measurement, entanglement, and the no-cloning theorem, all features with no classical counterparts. These characteristics open the possibility of designing learning models with novel forms of resilience and security. Quantum cryptographic techniques such as quantum key distribution already demonstrate how quantum principles can secure classical communication, for example by using measurement collapse to detect eavesdropping.

In a similar spirit, QML models may exhibit forms of robustness inherently tied to the physics of quantum systems \cite{gao2022enhancing}. This inherent property of QML systems, can potentially offering better defenses against specific classes of adversarial attacks. At the same time, as QML continues to advance and potentially find more relevance in applications, it becomes increasingly important to examine its robustness in adversarial settings \cite{liu2020vulnerability}. The very quantum mechanical properties that offer the expected advantages may also introduce new kinds of vulnerabilities. These susceptibilities can emerge from sensitivity to noise and decoherence to the possibility of adversaries exploiting quantum interference or circuit design. In this light, QML represents both an opportunity and a challenge. It offers promising new tools to address adversarial threats in machine learning, but it also demands new theoretical and practical frameworks to understand its behavior under worst-case conditions. Studying the robustness of quantum learning systems is, therefore, essential to the responsible development of the field.

Throughout this book, robustness often refers to adversarial robustness. This is defined to the ability of quantum machine learning models to maintain performance in the presence of carefully crafted, maliciously chosen inputs designed to cause system digressions or failure. In the QML context, these adversarial examples may exploit flaws in data encoding, circuit design, or optimization routines \cite{west2024drastic}. This notion of robustness can be clearly distinguished from other uses of the term, such as in optimization theory which manifests as the sensitivity to perturbations in cost functions, or statistical learning where robustness refers to that towards outliers or distributions.

If we imagine plotting the role of the adversary in QML systems, it can be captured by adding a third dimension to the widely used data-device framework, as illustrated in Fig. \ref{fig:dim3robustness}. Traditionally, QML studies focus on whether the model and data are classical or quantum, often exploring the advantages of quantum models acting on classical or quantum data. One can extend the third axes to conceptual analysis of the adversary in each relevant scenario.

By doing so, we add a new layer of analysis. Each quadrant in this cube represents a data-device-adversary pairing, capturing the nature and capabilities of the adversary as the novel extension. For instance, an adversary restricted to classical computation may be unable to efficiently craft perturbations against quantum data encoded in high-dimensional Hilbert spaces, while a quantum adversary might exploit entanglement or interference to mount entirely new types of attacks. Similarly, a classical device operating on quantum data may face vulnerabilities in the measurement interface, whereas a quantum device processing classical data may be more exposed through its optimization routines. 

Viewing QML robustness in this three-dimensional space re-frames familiar questions. Instead of simply asking \textit{``What advantage does a quantum model offer on classical data?''}, one can ask \textit{``Does this advantage persist under the capabilities of a quantum adversary, and if not, why?''} This cubic framework allows us to systematically examine the attack surfaces that exist in each region and analyze how the adversary’s computational model interacts with the data and device to shape vulnerability.

In this sense, this conceptual framework underlines that quantum computing may offer new tools to address robustness challenges, either through fundamentally different model behavior or through novel algorithmic defenses. These new computational capabilities as well as new sources of fragility define the next stage of research in the field. The next chapters will cover more in-depth the vulnerabilities quantum phenomena introduce into learning, as well as the techniques used to mitigate these vulnerabilities and enhance model robustness.

\section{Conclusion and Outlook}
This chapter has outlined the motivations, foundations, and interdisciplinary nature of quantum machine learning. QML is propelled by the successes and limitations of classical machine learning, and it is grounded on the quantum computational paradigm that is still being actively shaped. As outlined in this chapter, this dual identity brings both promise and uncertainty. The promise comes from the opportunities quantum structures presents. As we elaborated above, one such promise is that of computational speedups, inherited to QML models via the fundamental linear algebra subroutines. Another important opportunity is brought forth by quantum computers expected ability to capture data representations in higher-dimensional Hilbert spaces, potentially allowing for greater pattern emergence and more expressive models.

However, its practical realization is hindered by today's constrained hardware capacities which suffers from noise, decoherence, and limited qubit connectivity. Many theoretical models assume operations that far exceed today's capabilities requiring the full processing power of fault-tolerant quantum computers. Furthermore, within the quantum-enhanced machine learning paradigm, data encoding as a pre-processing step is a resource-intensive process. Optimization of quantum and hybrid learning models remains a major bottleneck, with variational algorithms commonly plagued by barren plateaus that hinder effective training. Moreover, the lack of standardized quantum-native datasets limits rigorous evaluation and understanding of what quantum phenomena can bring to the learning process.

Nevertheless, research in QML remains active and determined, driven by long-term technological promise, interdisciplinary momentum, and theoretical curiosity. While significant obstacles persist, these very challenges have spurred a rich body of work aimed at both understanding and overcoming them. On an algorithmic level, a growing line of work focuses on understanding and mitigating barren plateaus via architectural design, initialization strategies, and local cost functions. These efforts are part of the broader initiative to develop trainable, scalable, and expressive quantum models, without losing sight of how they measure up to classical techniques. At a more fundamental level, an existing avenue explores theoretical bounds that characterize what quantum models can and cannot learn efficiently, again, often compared to their classical counterparts. These characterizations range from sample complexity to expressive power and quantum generalization bounds. Meanwhile, across the broader landscape of quantum computing there is an active push for error correction, error mitigation and coherent control strategies at scale. These foundational advancements are expected to benefit a wide range of broader quantum applications, including those enabled or advanced by QML.

Given the practical challenges posed by quantum hardware and algorithmic instability, QML systems demand careful benchmarking and thorough exploration. One important aspect of this evaluation is robustness: the capacity of a QML model to perform reliably in the presence of noise, or intentional perturbations. And because quantum systems are inherently sensitive to environmental factors such as decoherence and noise, this fragility makes the study of robustness especially relevant. Among the various dimensions of robustness, adversarial robustness is of particular interest. We hope this chapter has equipped readers with a foundational understanding of QML as a rapidly evolving interdisciplinary field, with a particular focus on the issue of robustness. In the chapters that follow, we will delve deeper into the specific vulnerabilities introduced by quantum phenomena, explore formal definitions and frameworks for robustness in QML, and examine current approaches to adversarial defense. As we move forward, the next chapters will dissect quantum-specific vulnerabilities, develop formal frameworks for robustness, and explore defenses that harness quantum principles rather than merely tolerate them. By the end of the book, readers will gain a comprehensive understanding of what it entails to design and develop robust QML systems.

\begin{acknowledgement}
We thank Debbie Lim and Josep Lumbreras for discussions. Portions of this manuscript were drafted or edited with the assistance of ChatGPT to improve clarity and style.
\end{acknowledgement}

\bibliographystyle{unsrt}
\bibliography{bibliography}

@article{aaronson2015read,
	title        = {Read the fine print},
	author       = {Aaronson, Scott},
	year         = {2015},
	journal      = {Nature Physics},
	publisher    = {Nature Publishing Group},
	volume       = {11},
	number       = {4},
	pages        = {291--293}
}

@article{arunachalam2017survey,
	title        = {Guest Column: A Survey of Quantum Learning Theory},
	author       = {Arunachalam, Srinivasan and de Wolf, Ronald},
	year         = {2017},
	month        = jun,
	journal      = {SIGACT News},
	publisher    = {Association for Computing Machinery},
	address      = {New York, NY, USA},
	volume       = {48},
	number       = {2},
	pages        = {41–67},
	doi          = {10.1145/3106700.3106710},
	issn         = {0163-5700},
	url          = {https://doi.org/10.1145/3106700.3106710},
	issue_date   = {June 2017},
	numpages     = {27}
}

@article{arute2019quantum,
	title        = {Quantum supremacy using a programmable superconducting processor},
	author       = {Frank Arute and Kunal Arya and Ryan Babbush and et al.},
	year         = {2019},
	journal      = {Nature},
	publisher    = {Nature Publishing Group},
	volume       = {574},
	pages        = {505--510},
	doi          = {10.1038/s41586-019-1666-5},
	url          = {https://doi.org/10.1038/s41586-019-1666-5}
}

@article{belkin2019reconciling,
	title        = {Reconciling Modern Machine-Learning Practice and the Classical Bias--Variance Trade-Off},
	author       = {Belkin, Mikhail and Hsu, Daniel and Ma, Siyuan and Mandal, Soumik},
	year         = {2019},
	journal      = {Proceedings of the National Academy of Sciences},
	volume       = {116},
	number       = {32},
	pages        = {15849--15854},
	doi          = {10.1073/pnas.1903070116}
}

@article{ben2010theory,
	title        = {A Theory of Learning from Different Domains},
	author       = {Ben-David, Shai and Blitzer, John and Crammer, Koby and Kulesza, Alex and Pereira, Fernando and Vaughan, Jennifer Wortman},
	year         = {2010},
	journal      = {Machine Learning},
	volume       = {79},
	number       = {1},
	pages        = {151--175},
	doi          = {10.1007/s10994-009-5152-4}
}

@book{ben2013robust,
	title        = {Robust Optimization},
	author       = {Ben-Tal, Aharon and El Ghaoui, Laurent and Nemirovski, Arkadi},
	year         = {2009},
	publisher    = {Princeton University Press},
	doi          = {10.1515/9781400831050}
}

@article{benedetti2019parameterized,
	title        = {Parameterized quantum circuits as machine learning models},
	author       = {Benedetti,  Marcello and Lloyd,  Erika and Sack,  Stefan and Fiorentini,  Mattia},
	year         = {2019},
	month        = nov,
	journal      = {Quantum Science and Technology},
	publisher    = {IOP Publishing},
	volume       = {4},
	number       = {4},
	pages        = {043001},
	doi          = {10.1088/2058-9565/ab4eb5},
	issn         = {2058-9565},
	url          = {http://doi.org/10.1088/2058-9565/ab4eb5}
}

@article{benioff1980computer,
	title        = {The computer as a physical system: A microscopic quantum mechanical Hamiltonian model of computers as represented by Turing machines},
	author       = {Benioff, Paul},
	year         = {1980},
	journal      = {Journal of Statistical Physics},
	publisher    = {Springer},
	volume       = {22},
	number       = {5},
	pages        = {563--591},
	doi          = {10.1007/BF01011339}
}

@article{biamonte2017quantum,
	title        = {Quantum machine learning},
	author       = {Biamonte, Jacob and Wittek, Peter and Pancotti, Nicola and Rebentrost, Patrick and Wiebe, Nathan and Lloyd, Seth},
	year         = {2017},
	month        = sep,
	journal      = {Nature},
	publisher    = {Springer Science and Business Media LLC},
	volume       = {549},
	number       = {7671},
	pages        = {195–202},
	doi          = {10.1038/nature23474},
	issn         = {1476-4687},
	url          = {http://dx.doi.org/10.1038/nature23474}
}

@book{bishop2006pattern,
	title        = {Pattern Recognition and Machine Learning},
	author       = {Bishop, Christopher M.},
	year         = {2006},
	publisher    = {Springer},
	isbn         = {978-0387310732}
}

@misc{bowles2024betterclassicalsubtleart,
	title        = {Better than classical? The subtle art of benchmarking quantum machine learning models},
	author       = {Joseph Bowles and Shahnawaz Ahmed and Maria Schuld},
	year         = {2024},
	url          = {https://arxiv.org/abs/2403.07059},
	eprint       = {2403.07059},
	archiveprefix = {arXiv},
	primaryclass = {quant-ph}
}

@inproceedings{brown2020language,
	title        = {Language Models are Few-Shot Learners},
	author       = {Brown, Tom and Mann, Benjamin and Ryder, Nick and et al.},
	year         = {2020},
	booktitle    = {Advances in Neural Information Processing Systems},
	eprint       = {2005.14165},
	archiveprefix = {arXiv},
	primaryclass = {cs.CL}
}

@article{cao2019quantum,
	title        = {Quantum Chemistry in the Age of Quantum Computing},
	author       = {Cao, Yudong and Romero, Jonathan and Olson, Jonathan P. and Degroote, Matthias and Johnson, Peter D. and Kieferová, Mária and Kivlichan, Ian D. and Menke, Timothy and Peropadre, Borja and Sawaya, Nicolas P. D. and Sim, Sukin and Veis, Libor and Aspuru-Guzik, Alán},
	year         = {2019},
	month        = oct,
	journal      = {Chemical Reviews},
	publisher    = {American Chemical Society},
	volume       = {119},
	number       = {19},
	pages        = {10856--10915},
	doi          = {10.1021/acs.chemrev.8b00803}
}

@article{caro2022generalizing,
	title        = {Generalization in quantum machine learning from few training data},
	author       = {Caro, Marco C. and Huang, Hsin-Yuan and Cerezo, M. and Sharma, Kunal and Sornborger, Andrew T. and Coles, Patrick J.},
	year         = {2022},
	journal      = {Nature Communications},
	volume       = {13},
	pages        = {4919},
	doi          = {10.1038/s41467-022-32550-3}
}

@article{cerezo2021variational,
	title        = {Variational Quantum Algorithms},
	author       = {Cerezo, M. and Arrasmith, A. and Babbush, R. and Benjamin, S. C. and Endo, S. and Fujii, K. and McClean, J. R. and Mitarai, K. and Yuan, X. and Cincio, Ł. and Coles, P. J.},
	year         = {2021},
	journal      = {Nature Reviews Physics},
	publisher    = {Springer Nature},
	volume       = {3},
	pages        = {625--644},
	doi          = {10.1038/s42254-021-00348-9},
	url          = {https://www.nature.com/articles/s42254-021-00348-9}
}

@article{cortes1995support,
	title        = {Support-Vector Networks},
	author       = {Cortes, Corinna and Vapnik, Vladimir},
	year         = {1995},
	journal      = {Machine Learning},
	volume       = {20},
	pages        = {273--297},
	doi          = {10.1007/BF00994018}
}

@inproceedings{dalvi2004adversarial,
	title        = {Adversarial Classification},
	author       = {Dalvi, Nilesh and Domingos, Pedro and Sanghai, Sumit and Verma, Deepak},
	year         = {2004},
	booktitle    = {ACM SIGKDD International Conference on Knowledge Discovery and Data Mining},
	pages        = {99--108},
	doi          = {10.1145/1014052.1014118}
}

@article{duchi2016statistics,
	title        = {Statistics of Robust Optimization: A Generalized Empirical Likelihood Approach},
	author       = {Duchi, John C. and Namkoong, Hongseok},
	year         = {2022},
	journal      = {Mathematics of Operations Research},
	volume       = {47},
	number       = {2},
	pages        = {882--918},
	doi          = {10.1287/moor.2021.1193},
	note         = {(Original preprint 2016)}
}

@article{farhi2014quantumapproximate,
	title        = {A Quantum Approximate Optimization Algorithm},
	author       = {Farhi, Edward and Goldstone, Jeffrey and Gutmann, Sam},
	year         = {2014},
	month        = nov,
	journal      = {arXiv preprint arXiv:1411.4028},
	url          = {https://arxiv.org/abs/1411.4028}
}

@article{feynman1982simulating,
	title        = {Simulating Physics with Computers},
	author       = {Feynman, Richard P.},
	year         = {1982},
	journal      = {International Journal of Theoretical Physics},
	volume       = {21},
	number       = {6-7},
	pages        = {467--488},
	doi          = {10.1007/BF02650179}
}

@article{freund1997decision,
	title        = {A Decision-Theoretic Generalization of On-Line Learning and an Application to Boosting},
	author       = {Freund, Yoav and Schapire, Robert E.},
	year         = {1997},
	journal      = {Journal of Computer and System Sciences},
	volume       = {55},
	number       = {1},
	pages        = {119--139},
	doi          = {10.1006/jcss.1997.1504}
}

@article{gao2022enhancing,
	title        = {Enhancing Generative Models via Quantum Correlations},
	author       = {Gao, Xun and Anschuetz, Eric R. and Wang, Sheng-Tao and Cirac, J. Ignacio and Lukin, Mikhail D.},
	year         = {2022},
	journal      = {Physical Review X},
	volume       = {12},
	number       = {2},
	pages        = {021037},
	doi          = {10.1103/PhysRevX.12.021037}
}

@inproceedings{gilyen2019quantum,
	title        = {Quantum singular value transformation and beyond: exponential improvements for quantum matrix arithmetics},
	author       = {Gilyén, András and Su, Yuan and Low, Guang Hao and Wiebe, Nathan},
	year         = {2019},
	month        = jun,
	booktitle    = {Proceedings of the 51st Annual ACM SIGACT Symposium on Theory of Computing},
	publisher    = {ACM},
	series       = {STOC ’19},
	pages        = {193–204},
	doi          = {10.1145/3313276.3316366},
	url          = {http://dx.doi.org/10.1145/3313276.3316366},
	collection   = {STOC ’19}
}

@book{goodfellow2016deep,
	title        = {Deep learning},
	author       = {Goodfellow, Ian and Bengio, Yoshua and Courville, Aaron},
	year         = {2016},
	publisher    = {MIT Press},
	address      = {Cambridge, MA},
	url          = {http://www.deeplearningbook.org}
}

@article{harrow2009quantum,
	title        = {Quantum Algorithm for Linear Systems of Equations},
	author       = {Harrow, Aram W. and Hassidim, Avinatan and Lloyd, Seth},
	year         = {2009},
	month        = oct,
	journal      = {Physical Review Letters},
	publisher    = {American Physical Society (APS)},
	volume       = {103},
	number       = {15},
	doi          = {10.1103/physrevlett.103.150502},
	issn         = {1079-7114},
	url          = {http://dx.doi.org/10.1103/PhysRevLett.103.150502}
}

@article{harrow2017quantum,
	title        = {Quantum computational supremacy},
	author       = {Aram W. Harrow and Ashley Montanaro},
	year         = {2017},
	journal      = {Nature},
	publisher    = {Nature Publishing Group},
	volume       = {549},
	number       = {7671},
	pages        = {203--209},
	doi          = {10.1038/nature23458},
	url          = {https://doi.org/10.1038/nature23458}
}

@book{hastie2009elements,
	title        = {The Elements of Statistical Learning},
	author       = {Hastie, Trevor and Tibshirani, Robert and Friedman, Jerome},
	year         = {2009},
	publisher    = {Springer},
	doi          = {10.1007/978-0-387-84858-7},
	edition      = {2}
}

@article{havlicek2019supervised,
	title        = {Supervised Learning with Quantum-Enhanced Feature Spaces},
	author       = {Havlíček, V. and Córcoles, A. D. and Temme, K. and Harrow, A. W. and Kandala, A. and Chow, J. M. and Gambetta, J. M.},
	year         = {2019},
	journal      = {Nature},
	publisher    = {Springer Nature},
	volume       = {567},
	pages        = {209--212},
	doi          = {10.1038/s41586-019-0980-2},
	url          = {https://www.nature.com/articles/s41586-019-0980-2}
}

@inproceedings{he2016deep,
	title        = {Deep Residual Learning for Image Recognition},
	author       = {He, Kaiming and Zhang, Xiangyu and Ren, Shaoqing and Sun, Jian},
	year         = {2016},
	booktitle    = {IEEE Conference on Computer Vision and Pattern Recognition (CVPR)},
	pages        = {770--778},
	doi          = {10.1109/CVPR.2016.90}
}

@article{hinton2006fast,
	title        = {A Fast Learning Algorithm for Deep Belief Nets},
	author       = {Hinton, Geoffrey E. and Osindero, Simon and Teh, Yee-Whye},
	year         = {2006},
	journal      = {Neural Computation},
	volume       = {18},
	number       = {7},
	pages        = {1527--1554},
	doi          = {10.1162/neco.2006.18.7.1527}
}

@inproceedings{ho2020denoising,
	title        = {Denoising Diffusion Probabilistic Models},
	author       = {Ho, Jonathan and Jain, Ajay and Abbeel, Pieter},
	year         = {2020},
	booktitle    = {Advances in Neural Information Processing Systems},
	eprint       = {2006.11239},
	archiveprefix = {arXiv},
	primaryclass = {cs.LG}
}

@article{huang2021power,
	title        = {Power of data in quantum machine learning},
	author       = {Huang, Hsin-Yuan and Broughton, Michael and Mohseni, Masoud and Babbush, Ryan and Boixo, Sergio and Neven, Hartmut and McClean, Jarrod R.},
	year         = {2021},
	month        = may,
	journal      = {Nature Communications},
	publisher    = {Springer Science and Business Media LLC},
	volume       = {12},
	number       = {1},
	doi          = {10.1038/s41467-021-22539-9},
	issn         = {2041-1723},
	url          = {http://dx.doi.org/10.1038/s41467-021-22539-9}
}

@article{huber1964robust,
	title        = {Robust Estimation of a Location Parameter},
	author       = {Huber, Peter J.},
	year         = {1964},
	journal      = {Annals of Mathematical Statistics},
	volume       = {35},
	number       = {1},
	pages        = {73--101},
	doi          = {10.1214/aoms/1177703732}
}

@inproceedings{ioffe2015batch,
	title        = {Batch Normalization: Accelerating Deep Network Training by Reducing Internal Covariate Shift},
	author       = {Ioffe, Sergey and Szegedy, Christian},
	year         = {2015},
	booktitle    = {International Conference on Machine Learning},
	eprint       = {1502.03167},
	archiveprefix = {arXiv},
	primaryclass = {cs.LG}
}

@article{ivashkov2024qkanquantumkolmogorovarnoldnetworks,
	title        = {QKAN: Quantum Kolmogorov-Arnold Networks},
	author       = {Petr Ivashkov and Po-Wei Huang and Kelvin Koor and Lirandë Pira and Patrick Rebentrost},
	year         = {2024},
	journal      = {arXiv preprint arXiv:2410.04435},
	url          = {https://arxiv.org/abs/2410.04435},
	eprint       = {2410.04435},
	archiveprefix = {arXiv},
	primaryclass = {quant-ph}
}

@article{kaplan2020scaling,
	title        = {Scaling Laws for Neural Language Models},
	author       = {Jared Kaplan and Sam McCandlish and Tom Henighan and Tom B. Brown and Benjamin Chess and Rewon Child and Scott Gray and Alec Radford and Jeffrey Wu and Dario Amodei},
	year         = {2020},
	journal      = {arXiv preprint arXiv:2001.08361},
	eprint       = {2001.08361},
	archiveprefix = {arXiv},
	primaryclass = {cs.LG}
}

@article{kingma2015adam,
	title        = {Adam: A Method for Stochastic Optimization},
	author       = {Kingma, Diederik P. and Ba, Jimmy},
	year         = {2015},
	journal      = {International Conference on Learning Representations},
	eprint       = {1412.6980},
	archiveprefix = {arXiv},
	primaryclass = {cs.LG}
}

@book{koller2009probabilistic,
	title        = {Probabilistic Graphical Models: Principles and Techniques},
	author       = {Koller, Daphne and Friedman, Nir},
	year         = {2009},
	publisher    = {MIT Press},
	doi          = {10.7551/mitpress/7483.001.0001}
}

@inproceedings{krizhevsky2012imagenet,
	title        = {ImageNet Classification with Deep Convolutional Neural Networks},
	author       = {Krizhevsky, Alex and Sutskever, Ilya and Hinton, Geoffrey E.},
	year         = {2012},
	booktitle    = {Advances in Neural Information Processing Systems},
	eprint       = {1409.1556},
	archiveprefix = {arXiv},
	primaryclass = {cs.CV}
}

@article{larocca2025barren,
	title        = {Barren plateaus in variational quantum computing},
	author       = {Larocca, Martín and Thanasilp, Supanut and Wang, Samson and Sharma, Kunal and Biamonte, Jacob and Coles, Patrick J. and Cincio, Lukasz and McClean, Jarrod R. and Holmes, Zoë and Cerezo, M.},
	year         = {2025},
	month        = mar,
	journal      = {Nature Reviews Physics},
	publisher    = {Springer Science and Business Media LLC},
	volume       = {7},
	number       = {4},
	pages        = {174–189},
	doi          = {10.1038/s42254-025-00813-9},
	issn         = {2522-5820},
	url          = {http://dx.doi.org/10.1038/s42254-025-00813-9}
}

@article{larose2020robust,
	title        = {Robust data encodings for quantum classifiers},
	author       = {LaRose, Ryan and Coyle, Brian},
	year         = {2020},
	month        = {Sep},
	journal      = {Phys. Rev. A},
	publisher    = {American Physical Society},
	volume       = {102},
	pages        = {032420},
	doi          = {10.1103/PhysRevA.102.032420},
	url          = {https://link.aps.org/doi/10.1103/PhysRevA.102.032420},
	issue        = {3},
	numpages     = {24}
}

@article{lecun2015deep,
	title        = {Deep learning},
	author       = {LeCun, Yann and Bengio, Yoshua and Hinton, Geoffrey},
	year         = {2015},
	journal      = {Nature},
	publisher    = {Nature Publishing Group},
	volume       = {521},
	number       = {7553},
	pages        = {436--444}
}

@article{liu2020vulnerability,
	title        = {Vulnerability of quantum classification to adversarial perturbations},
	author       = {Liu, Nana and Wittek, Peter},
	year         = {2020},
	month        = {Jun},
	journal      = {Phys. Rev. A},
	publisher    = {American Physical Society},
	volume       = {101},
	pages        = {062331},
	doi          = {10.1103/PhysRevA.101.062331},
	url          = {https://link.aps.org/doi/10.1103/PhysRevA.101.062331},
	issue        = {6},
	numpages     = {9}
}

@article{lloyd2014quantum,
	title        = {Quantum principal component analysis},
	author       = {Lloyd, Seth and Mohseni, Masoud and Rebentrost, Patrick},
	year         = {2014},
	journal      = {Nature Physics},
	publisher    = {Nature Publishing Group},
	volume       = {10},
	number       = {9},
	pages        = {631--633},
	doi          = {10.1038/nphys3029}
}

@inproceedings{lowd2005adversarial,
	title        = {Adversarial Learning},
	author       = {Lowd, Daniel and Meek, Christopher},
	year         = {2005},
	booktitle    = {ACM SIGKDD International Conference on Knowledge Discovery and Data Mining},
	pages        = {641--647},
	doi          = {10.1145/1081870.1081950}
}

@article{lu2020quantum,
	title        = {Quantum adversarial machine learning},
	author       = {Lu, Sirui and Duan, Lu-Ming and Deng, Dong-Ling},
	year         = {2020},
	month        = {Aug},
	journal      = {Phys. Rev. Res.},
	publisher    = {American Physical Society},
	volume       = {2},
	pages        = {033212},
	doi          = {10.1103/PhysRevResearch.2.033212},
	url          = {https://link.aps.org/doi/10.1103/PhysRevResearch.2.033212},
	issue        = {3},
	numpages     = {22}
}

@article{mcclean2018barren,
	title        = {Barren plateaus in quantum neural network training landscapes},
	author       = {McClean, Jarrod R and Boixo, Sergio and Smelyanskiy, Vadim N and Babbush, Ryan and Neven, Hartmut},
	year         = {2018},
	journal      = {Nature Communications},
	publisher    = {Nature Publishing Group},
	volume       = {9},
	number       = {1},
	pages        = {4812},
	doi          = {10.1038/s41467-018-07090-4},
	url          = {https://doi.org/10.1038/s41467-018-07090-4}
}

@article{mitarai2018quantum,
	title        = {Quantum circuit learning},
	author       = {Mitarai, K. and Negoro, M. and Kitagawa, M. and Fujii, K.},
	year         = {2018},
	month        = sep,
	journal      = {Phys. Rev. A},
	publisher    = {American Physical Society},
	volume       = {98},
	pages        = {032309},
	url          = {https://doi.org/10.1103/PhysRevA.98.032309},
	issue        = {3},
	numpages     = {6}
}

@article{montanaro2016quantum,
	title        = {Quantum algorithms: an overview},
	author       = {Montanaro, Ashley},
	year         = {2016},
	month        = jan,
	journal      = {npj Quantum Information},
	publisher    = {Springer Science and Business Media LLC},
	volume       = {2},
	number       = {1},
	doi          = {10.1038/npjqi.2015.23},
	issn         = {2056-6387},
	url          = {http://dx.doi.org/10.1038/npjqi.2015.23}
}

@misc{morales2025quantumlinearsolverssurvey,
	title        = {Quantum Linear System Solvers: A Survey of Algorithms and Applications},
	author       = {Mauro E. S. Morales and Lirandë Pira and Philipp Schleich and Kelvin Koor and Pedro C. S. Costa and Dong An and Alán Aspuru-Guzik and Lin Lin and Patrick Rebentrost and Dominic W. Berry},
	year         = {2025},
	url          = {https://arxiv.org/abs/2411.02522},
	eprint       = {2411.02522},
	archiveprefix = {arXiv},
	primaryclass = {quant-ph}
}

@inproceedings{namkoong2017variance,
	title        = {Variance-Based Regularization with Convex Objectives},
	author       = {Namkoong, Hongseok and Duchi, John C.},
	year         = {2017},
	booktitle    = {Advances in Neural Information Processing Systems},
	eprint       = {1610.02581},
	archiveprefix = {arXiv},
	primaryclass = {stat.ML}
}

@article{nesterov1983method,
	title        = {A Method of Solving a Convex Programming Problem with Convergence Rate $O(1/k^2)$},
	author       = {Nesterov, Yurii},
	year         = {1983},
	journal      = {Soviet Mathematics Doklady},
	volume       = {27},
	pages        = {372--376}
}

@book{nielsen2010quantum,
	title        = {Quantum Computation and Quantum Information},
	author       = {Michael A. Nielsen and Isaac L. Chuang},
	year         = {2010},
	publisher    = {Cambridge University Press},
	address      = {Cambridge, UK},
	isbn         = {978-1107002173},
	edition      = {10th Anniversary Edition}
}

@article{orus2019quantum,
	title        = {Quantum computing for finance: Overview and prospects},
	author       = {Orús, Román and Mugel, Samuel and Lizaso, Enrique},
	year         = {2019},
	month        = nov,
	journal      = {Reviews in Physics},
	publisher    = {Elsevier BV},
	volume       = {4},
	pages        = {100028},
	doi          = {10.1016/j.revip.2019.100028},
	issn         = {2405-4283},
	url          = {http://dx.doi.org/10.1016/j.revip.2019.100028}
}

@article{perezsalinas2020data,
	title        = {Data re-uploading for a universal quantum classifier},
	author       = {P{\'e}rez-Salinas, Adri{\'a} and Cervera-Lierta, Alberto and Gil-Fuster, Elies and Latorre, Jos{\'e} I.},
	year         = {2020},
	journal      = {Quantum},
	volume       = {4},
	pages        = {226},
	doi          = {10.22331/q-2020-02-06-226},
	url          = {https://quantum-journal.org/papers/q-2020-02-06-226/}
}

@article{peruzzo2014variational,
	title        = {A Variational Eigenvalue Solver on a Photonic Quantum Processor},
	author       = {Peruzzo, Alberto and McClean, Jarrod and Shadbolt, Peter and Yung, Man-Hong and Zhou, Xiao-Qi and Love, Peter J. and Aspuru-Guzik, Alán and O'Brien, Jeremy L.},
	year         = {2014},
	journal      = {Nature Communications},
	publisher    = {Nature Publishing Group},
	volume       = {5},
	pages        = {4213},
	doi          = {10.1038/ncomms5213}
}

@article{preskill2018quantum,
	title        = {Quantum computing in the {NISQ} era and beyond},
	author       = {Preskill, John},
	year         = {2018},
	journal      = {Quantum},
	volume       = {2},
	pages        = {79},
	doi          = {10.22331/q-2018-08-06-79}
}

@article{rebentrost2014quantum,
	title        = {Quantum Support Vector Machine for Big Data Classification},
	author       = {Rebentrost, Patrick and Mohseni, Masoud and Lloyd, Seth},
	year         = {2014},
	month        = sep,
	journal      = {Physical Review Letters},
	publisher    = {American Physical Society (APS)},
	volume       = {113},
	number       = {13},
	doi          = {10.1103/physrevlett.113.130503},
	issn         = {1079-7114},
	url          = {http://dx.doi.org/10.1103/PhysRevLett.113.130503}
}

@article{ren2022experimental,
	title        = {Experimental quantum adversarial learning with programmable superconducting qubits},
	author       = {Ren, W. and Li, W. and Xu, S. and et al.},
	year         = {2022},
	journal      = {Nature Computational Science},
	publisher    = {Nature Publishing Group},
	volume       = {2},
	pages        = {711--717},
	doi          = {10.1038/s43588-022-00351-9}
}

@article{robbins1951stochastic,
	title        = {A Stochastic Approximation Method},
	author       = {Robbins, Herbert and Monro, Sutton},
	year         = {1951},
	journal      = {Annals of Mathematical Statistics},
	volume       = {22},
	number       = {3},
	pages        = {400--407},
	doi          = {10.1214/aoms/1177729586}
}

@article{rosenblatt1958perceptron,
	title        = {The Perceptron: A Probabilistic Model for Information Storage and Organization in the Brain},
	author       = {Rosenblatt, Frank},
	year         = {1958},
	journal      = {Psychological Review},
	volume       = {65},
	number       = {6},
	pages        = {386--408},
	doi          = {10.1037/h0042519}
}

@article{rumelhart1986learning,
	title        = {Learning Representations by Back-Propagating Errors},
	author       = {Rumelhart, David E. and Hinton, Geoffrey E. and Williams, Ronald J.},
	year         = {1986},
	journal      = {Nature},
	volume       = {323},
	pages        = {533--536},
	doi          = {10.1038/323533a0}
}

@book{scholkopf2002learning,
	title        = {Learning with Kernels},
	author       = {Sch{\"o}lkopf, Bernhard and Smola, Alexander J.},
	year         = {2002},
	publisher    = {MIT Press},
	doi          = {10.7551/mitpress/4175.001.0001}
}

@article{schuld2019evaluating,
	title        = {Evaluating analytic gradients on quantum hardware},
	author       = {Schuld, Maria and Bergholm, Ville and Gogolin, Christian and Izaac, Josh and Killoran, Nathan},
	year         = {2019},
	month        = mar,
	journal      = {Physical Review A},
	publisher    = {American Physical Society (APS)},
	volume       = {99},
	number       = {3},
	doi          = {10.1103/physreva.99.032331}
}

@article{schuld2019quantum,
	title        = {Quantum Machine Learning in Feature Hilbert Spaces},
	author       = {Schuld, Maria and Killoran, Nathan},
	year         = {2019},
	month        = feb,
	journal      = {Physical Review Letters},
	publisher    = {American Physical Society (APS)},
	volume       = {122},
	number       = {4},
	doi          = {10.1103/physrevlett.122.040504},
	issn         = {1079-7114},
	url          = {http://dx.doi.org/10.1103/PhysRevLett.122.040504}
}

@book{schuld2021machine,
	title        = {Machine Learning with Quantum Computers},
	author       = {Schuld, Maria and Petruccione, Francesco},
	year         = {2021},
	publisher    = {Springer},
	address      = {Cham, Switzerland},
	series       = {Quantum Science and Technology},
	doi          = {10.1007/978-3-030-83098-4},
	isbn         = {978-3‑030‑83098‑4},
	edition      = {2}
}

@misc{shor1997faulttolerantquantumcomputation,
	title        = {Fault-tolerant quantum computation},
	author       = {Peter W. Shor},
	year         = {1997},
	url          = {https://arxiv.org/abs/quant-ph/9605011},
	eprint       = {quant-ph/9605011},
	archiveprefix = {arXiv},
	primaryclass = {quant-ph}
}

@inproceedings{shor1997polynomial,
	title        = {Polynomial-Time Algorithms for Prime Factorization and Discrete Logarithms on a Quantum Computer},
	author       = {Shor, Peter W.},
	year         = {1997},
	booktitle    = {SIAM Journal on Computing},
	publisher    = {Society for Industrial and Applied Mathematics},
	volume       = {26},
	pages        = {1484--1509},
	doi          = {10.1137/S0097539795293172}
}

@inproceedings{tang2019quantum,
	title        = {A quantum-inspired classical algorithm for recommendation systems},
	author       = {Tang, Ewin},
	year         = {2019},
	month        = jun,
	booktitle    = {Proceedings of the 51st Annual ACM SIGACT Symposium on Theory of Computing},
	publisher    = {ACM},
	series       = {STOC ’19},
	pages        = {217–228},
	doi          = {10.1145/3313276.3316310},
	url          = {http://dx.doi.org/10.1145/3313276.3316310},
	collection   = {STOC ’19}
}

@incollection{tukey1960survey,
	title        = {A Survey of Sampling from Contaminated Distributions},
	author       = {Tukey, John W.},
	year         = {1960},
	booktitle    = {Contributions to Probability and Statistics},
	publisher    = {Stanford University Press},
	pages        = {448--485}
}

@article{valiant1984theory,
	title        = {A theory of the learnable},
	author       = {Valiant, Leslie G.},
	year         = {1984},
	journal      = {Communications of the ACM},
	publisher    = {ACM},
	volume       = {27},
	number       = {11},
	pages        = {1134--1142},
	doi          = {10.1145/1968.1972},
	url          = {https://doi.org/10.1145/1968.1972}
}

@article{vapnik1971vc,
	title        = {On the Uniform Convergence of Relative Frequencies of Events to Their Probabilities},
	author       = {Vapnik, Vladimir and Chervonenkis, Alexey},
	year         = {1971},
	journal      = {Theory of Probability \& Its Applications},
	volume       = {16},
	number       = {2},
	pages        = {264--280},
	doi          = {10.1137/1116025}
}

@inproceedings{vaswani2017attention,
	title        = {Attention Is All You Need},
	author       = {Vaswani, Ashish and Shazeer, Noam and Parmar, Niki and Uszkoreit, Jakob and Jones, Llion and Gomez, Aidan N. and Kaiser, {\L}ukasz and Polosukhin, Illia},
	year         = {2017},
	booktitle    = {Advances in Neural Information Processing Systems},
	eprint       = {1706.03762},
	archiveprefix = {arXiv},
	primaryclass = {cs.CL}
}

@article{west2023towards,
	title        = {Towards quantum enhanced adversarial robustness in machine learning},
	author       = {West, M. T. and Tsang, S. L. and Low, J. S. and et al.},
	year         = {2023},
	journal      = {Nature Machine Intelligence},
	volume       = {5},
	pages        = {581--589},
	doi          = {10.1038/s42256-023-00661-1},
	url          = {https://doi.org/10.1038/s42256-023-00661-1}
}

@article{west2024drastic,
	title        = {Drastic Circuit Depth Reductions with Preserved Adversarial Robustness by Approximate Encoding for Quantum Machine Learning},
	author       = {West, Maxwell T. and Nakhl, Azar C. and Heredge, Jamie and Creevey, Floyd M. and Hollenberg, Lloyd C. L. and Sevior, Martin and Usman, Muhammad},
	year         = {2024},
	month        = jan,
	journal      = {Intelligent Computing},
	publisher    = {American Association for the Advancement of Science (AAAS)},
	volume       = {3},
	doi          = {10.34133/icomputing.0100},
	issn         = {2771-5892},
	url          = {http://dx.doi.org/10.34133/icomputing.0100}
}

@article{zhong2020quantum,
	title        = {Quantum computational advantage using photons},
	author       = {Han-Sen Zhong  and Hui Wang  and Yu-Hao Deng  and Ming-Cheng Chen  and Li-Chao Peng  and Yi-Han Luo  and Jian Qin  and Dian Wu  and Xing Ding  and Yi Hu  and Peng Hu  and Xiao-Yan Yang  and Wei-Jun Zhang  and Hao Li  and Yuxuan Li  and Xiao Jiang  and Lin Gan  and Guangwen Yang  and Lixing You  and Zhen Wang  and Li Li  and Nai-Le Liu  and Chao-Yang Lu  and Jian-Wei Pan},
	year         = {2020},
	journal      = {Science},
	volume       = {370},
	number       = {6523},
	pages        = {1460--1463},
	doi          = {10.1126/science.abe8770}
}

\end{document}